\documentclass[pdflatex, twocolumn]{sn-jnl}

\usepackage{graphicx}
\usepackage{amsmath,amssymb,amsfonts}

\usepackage{siunitx} 
\usepackage[braket, qm]{qcircuit} 

\usepackage[%
  backend=biber,
  url=true,
  style=numeric,
  sorting=none, 
  maxnames=3,
  minnames=1,
  maxbibnames=99,
  giveninits,
  uniquename=init]{biblatex} 
\bibliography{bibliography}

\usepackage[shortcuts,acronym]{glossaries} 
\setkeys{glslink}{hyper=false} 
\newacronym{lsb}{LSB}{Least Significant Bit}
\newacronym{msb}{MSB}{Most Significant Bit}
\newacronym{nisq}{NISQ}{Noisy Intermediate-Scale Quantum}
\newacronym{qfb}{QFB}{Quantum Feedback}
\newacronym{rca}{RCA}{Ripple-Carry-Adder}
\newacronym{rcu}{RCU}{Ripple-Carry-Unadder}

\usepackage{hyperref}

\newcommand{\cf}{\textrm{cf.}~} 
\newcommand{\eg}{\textrm{e.g.}~} 
\newcommand{\ie}{\textrm{i.e.}~} 
\newcommand{\Ie}{\textrm{I.e.}~} 

\newcommand{\apriori}{\textrm{a priori}~}

\begin{document}

\newcommand{\todo}[1]{\textbf{\textsc{\textcolor{red}{(TODO: #1)}}}} 

\title[Integer Factoring with Unoperations]{Integer Factoring with Unoperations}

\author[1]{\fnm{Paul} \sur{Kohl}}\email{paul.kohl@tum.de}

\affil[1]{\orgdiv{School of Computation, Information and Technology}, \orgname{Technical University of Munich}, \orgaddress{\street{Theresienstr. 90}, \postcode{80333} \city{Munich}, \state{Bavaria}, \country{Germany}}}

\abstract{This work introduces the notion of unoperation $\mathfrak{Un}(\hat{O})$ of some operation $\hat{O}$. Given a valid output of $\hat{O}$, the corresponding unoperation produces a set of all valid inputs to $\hat{O}$ that produce the given output. Further, the working principle of unoperations is illustrated using the example of addition. A device providing that functionality is constructed utilising a quantum circuit performing the unoperation of addition -- referred to as unaddition. To highlight the potential of the approach the unaddition quantum circuit is employed to construct a device for factoring integer numbers $N$, which is then called unmultiplier. This approach requires only a number of qubits $\in \mathcal{O}((\log{N})^2)$, rivalling the best known factoring algorithms to date.}

\keywords{Quantum Computing, Quantum Algorithms, Factoring, Digital Circuits}

\pacs[MSC Classification]{11A51, 68Q12, 81P65, 81P68}

\makeatletter
\twocolumn[\@maketitle]
\makeatother

\section{Introduction}\label{sec:Introduction}

In the past, it seemingly was hard to construct quantum algorithms for general problems. Because of this, the best known algorithms in quantum computing are probably still Grover's search algorithm \cite{GroversAlgorithm} and Shor's factoring algorithm \cite{ShorsAlgorithm, ShorsAlgorithm94}. Nevertheless, other approaches trying to utilise quantum resources via variational approaches have been explored and can even be tested on physical devices already \cite{VariationalQuantumEigensolver}. 
But none with the same critical impact as the aforementioned algorithms, as those variational approaches tend to be intended for \gls{nisq} devices.
Some other applications of quantum technology include the well established quantum key distribution protocols based on superposition \cite{BB84Protocol} or on entanglement \cite{E91Protocol}. There also exist proposals for quantum (identity) authentication protocols \cite{QuantumAuthShi, QuantumIdentityAuthHong, QuantumIdentityAuthZawadzki} and quantum message authentication codes \cite{QuantumMACCurty}, but some of them have weaknesses just as classical approaches \cite{CommentOnQuantumIdentityAuthenticationWithSinglePhoton, QuantumIdentityAuthZawadzki, QuantumAuthShiCommentWei, QuantumIdentityAuthZawadzkiAttackGonzalez}. This demonstrates that there are certain fields that might take advantage of the exploitation of quantum mechanical principles. Yet, a general framework to construct quantum algorithms that have an advantage over classical algorithms for different problems remains elusive.

One group of problems that would be of interest for novel algorithms in practice is the following. Problems that have an efficient classical algorithm, however, the inverse of the problem does not have any efficient algorithm (yet).
In the context of cryptography such a function is called a trapdoor one-way function \cite{DiffieHellman}.
If one would be able to construct a device that can efficiently reverse any efficiently computable operation, this hypothetical device would obviously open up the possibility of constructing a computationally easy inversion of such trapdoor functions. 

The most famous example for a trapdoor function is of course multiplication and factoring. RSA, still one of the most widely used public-key cryptography schemes is based on the computational hardness of factoring \cite{RSA}. For this trapdoor function Shor's algorithm provides an efficient algorithm on a quantum computer, thus this problem is a good candidate for comparison with the main contribution of this work, which aims at the construction of a quantum algorithm based on the concept of unoperations. This novel approach might enable a more systematic approach to constructing quantum algorithms for a certain class of problems, making it interesting for the whole field of quantum computing.

\section{Methods}\label{sec:Methods}

The striking fact is, that quantum circuits are inherently reversible due to the unitarity of quantum mechanical transformations. Therefore, in quantum computation the inversion of a computation on a quantum state to get back the original quantum state is called uncomputation.
However, this implies that there is already a quantum state produced by a preceding computation which can be uncomputed. In general, one is not in the fortunate situation to have an intact quantum state produced by the algorithm to be inverted.

\subsection{Unoperations}\label{subsec:Unoperations}

Because a readily uncomputable state is rarely the starting point I propose the notion of unoperation $\mathfrak{Un}(\cdot)$. That is, the so-called unoperation $\mathfrak{Un}(\hat{O})$ of some operation $\hat{O}$, where 
\begin{equation}
    \hat{O}(x) := \Tilde{x}
\end{equation}

is an operation on input $x$ producing output $\Tilde{x}$. 
Given a valid output $\Tilde{x}$ of operation $\hat{O}$ the unoperation is then defined as the function $\mathfrak{Un}(\hat{O})$ that produces a set of all valid inputs $x$ to $\hat{O}$ that produces the given output $\Tilde{x}$.
At the first glance, the proposed notion is not really that different from inversion of a function or uncomputation of a quantum state. At a second glance however, there is a subtle difference.
The unoperation coincides with the inverse of a function in case the function is bijective, because then there is a well-defined unique pairing of any input with its (single) corresponding output. Reformulated in the framework of quantum mechanics that corresponds to the statement that they coincide if the operator is unitary and thus has a conjugate transpose that can be applied to the state to uncompute it.
Now, the unoperation is not just limited to this case. The unoperation $\mathfrak{Un}(\hat{O})$ should produce \textit{all} valid inputs $x$ to $\hat{O}$ given $\Tilde{x}$.
That is, it is defined as
\begin{equation}
    \mathfrak{Un}(\hat{O})(\Tilde{x}) := \{x \:|\: \hat{O}(x) = \Tilde{x} \}.
\end{equation}

This means, we want to get a set of inputs and not just a single input producing $\Tilde{x}$, which can be easily illustrated for a surjective operation like addition.

\subsection{Unaddition}

Accordingly, we now turn to the unoperation of addition -- consequently called unaddition and denoted $\mathfrak{Un}{+}$ as the unoperation of the infix operator $+$ for addition.
Addition is a surjective operation and also a binary operation. As such it has two inputs and one output. We concentrate only on sums on field $\mathbb{N}_0$ here.
For example, if the output of the addition is $\Tilde{x}=3$, then the unaddition should produce
\begin{equation}
    \mathfrak{Un}{+}(3) = \{(0,3), (1,2), (2,1), (3,0)\}.
\end{equation}

\subsection{Construction of quantum circuits}\label{subsec:ConstructionQuantumCircuits}

To construct a quantum circuit that achieves this functionality, we first impose the following:
The result of the quantum mechanical unaddition operator should be a superposition of all the elements in the set that should be produced by unaddition. 
The device will be constructed utilising principles from classical circuit design adapted to the intended devices and taking into account capabilities of quantum circuits.
The results of the construction of the different quantum circuits, quantum gates and a full quantum algorithm for integer factoring will be given in Sec.~\ref{sec:Results} and discussed in Sec.~\ref{sec:Discussion}.

\subsection{Simulation of quantum circuits}

The devices were simulated with the classical quantum circuit simulator framework from \texttt{qiskit} \cite{Qiskit, QiskitSoftwarePackage} on a compute server with \qty{64}{} CPU threads at \qty{3,2}{\giga\hertz} maximum boost clock and \qty{128}{\gibi\byte} of DDR4-3200 RAM.

\section{Results}\label{sec:Results}

The device for unaddition, the so-called unadder was constructed in analogy to a classical digital circuit for addition.
The following section clarifies the concept of this digital circuit.

\subsection{Classical Ripple-Carry-Addition}

The \gls{rca} construction of the digital adder is a divide-and-conquer approach that breaks down the $n\:$\qty{}{\bit} numbers $a$ and $b$ to be added into single bits $a_i$ and $b_i$ with $i \in [0,n[$. These are then added with a full-adder device for single bit addition with carry. 
\cite{TechnischeInformatik1GrundlagenDerDigitalenElektronik, IntroLogicCircuits-ArithmeticCircuits}

Each full-adder performs following calculation:
\begin{equation}\label{eq:Full-Adder}
    a_i + b_i + c_{\text{in},i} = sum_i + c_{\text{out},i},
\end{equation}

where $a_i$, $b_i$, $c_{\text{in},i}$, $sum_i$, and $c_{\text{out},i}$ are single bits \cite[pp.188-189]{TechnischeInformatik1GrundlagenDerDigitalenElektronik}.
For the \gls{rca}, the carry $c_{\text{in},n-1} = 0$ for the bitwise addition of the \gls{lsb}. Additionally, the carry bit $c_{\text{out},i}$ of the addition of a less significant bit is fed into the next bitwise addition as the new $c_{\text{in},i-1}$. Like this the sum in the end can be constructed from the resulting bits $sum_i$ in the correct order and the final carry $c_{\text{out},0}$ is the \gls{msb} of the resulting sum.
\cite{TechnischeInformatik1GrundlagenDerDigitalenElektronik, IntroLogicCircuits-ArithmeticCircuits}

\subsection{Ripple-Carry-Unadder}

\begin{figure}[!bt]
    \centering
    \includegraphics[width=0.45\textwidth]{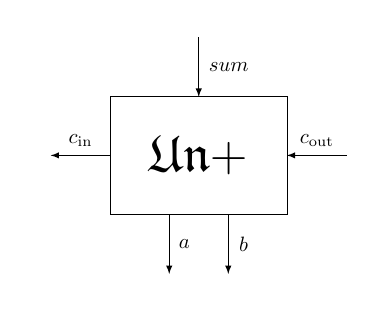}
    \caption[Black-box circuit of a full-unadder]{Black box circuit of a full-unadder. It is the unoperation device corresponding to the full-adder used in classical circuit design construction for \glspl{rca}. It is capable of the unaddition of a \qty{1}{\bit} sum with the carry $c_{\text{out}}$, which will produce valid bit combinations for $a$, $b$, and $c_{\text{in}}$ satisfying Eq.~\ref{eq:Full-Adder}.}
    \label{fig:Full-UnadderBlackBox}
\end{figure}

The analogous \gls{rcu} circuit utilises the same overall circuit design, but reverses the flow of information. To be able to do this, one needs to construct the unoperation of the bitwise unaddition, consequentially termed full-unadder. Its black-box circuit is shown in Fig.~\ref{fig:Full-UnadderBlackBox}. 

The \gls{rcu} then takes as input the $n\:$\qty{}{\bit} $sum$ in binary, such that its bits are $sum_i$ with $i \in [0,n[$. The final carry $c_{\text{out},0}$ -- which is an input now -- is assumed to be $0$. The output then should be the set of all $n\:$\qty{}{\bit} numbers $a$, $b$, and the carry bit $c_{\text{in},n-1}$ that satisfy
\begin{equation}\label{eq:RCU}
    a + b + c_{\text{in},n-1} = sum.
\end{equation}

In its quantum form the \gls{rcu} should produce the superposition of all the different elements of the set. Thus, I impose on the quantum full-unadder that it produces a uniform distribution of the valid output combinations in case an input of $(c_{\text{out},i}, sum_i)$ can be produced by multiple combinations. This is a reasonable assumption, because \apriori there is no knowledge that would suggest favouring one combination over the other. In consequence, the quantum operator of the full-unadder should satisfy the truth table given in Tab.~\ref{tab:Full-UnadderTruthTable}.

\begin{table}[ht]
    \caption[Truth table of a full-unadder]{Truth table of a full-unadder.}
    \label{tab:Full-UnadderTruthTable}
    \centering
    \begin{tabular}{  c  c  c  c  c  c }
        \toprule
            \multicolumn{2}{c}{Input} & \multicolumn{3}{c}{Output}  & Probability \\
            $c_{\text{out}}$  & $sum$ & $c_{\text{in}}$ & $b$ & $a$ & \\
        \midrule
            $0$               & $0$   & $0$ & $0$ & $0$ & $1$ \\
        \midrule
                              &       & $1$ & $0$ & $0$ & $1/3$ \\
            $0$               & $1$   & $0$ & $1$ & $0$ & $1/3$ \\
                              &       & $0$ & $0$ & $1$ & $1/3$ \\
        \midrule
                              &       & $0$ & $1$ & $1$ & $1/3$ \\
            $1$               & $0$   & $1$ & $0$ & $1$ & $1/3$ \\
                              &       & $1$ & $1$ & $0$ & $1/3$ \\
        \midrule
            $1$               & $1$   & $1$ & $1$ & $1$ & $1$ \\
        \botrule
    \end{tabular}
    \footnotetext{The full-unadder acts on single bits $sum$ and $c_{\text{out}}$ to produce the three bits $a$, $b$, and $c_{\text{in}}$ or a superposition of said bits in case the pair $(c_{\text{out}}, sum)$ can be produced by addition of different combinations of these bits. The probabilities for the output states are given in the last column for the given combination.}
\end{table}

\subsubsection{Quantum circuit for full-unadders}

\begin{figure*}[htb]
    \centering
    \includegraphics[width=1.0\textwidth]{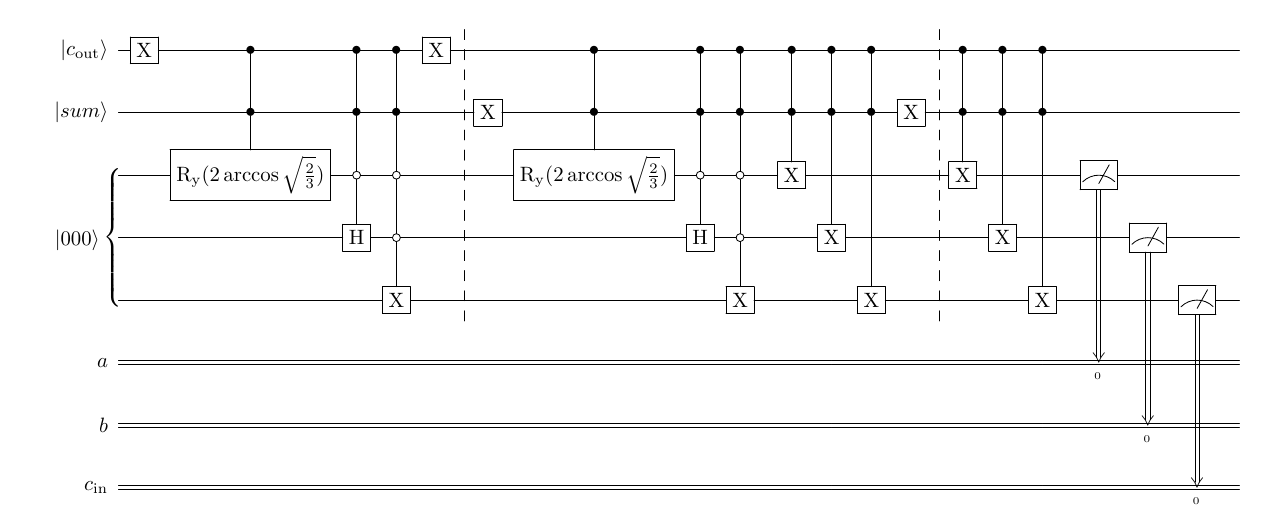}
    \caption[Quantum circuit of a full-unadder]{A quantum circuit of a full-unadder. The action on $\ket{c_{\text{out}}}\ket{sum} = \ket{00}$ is implicit and the actions on $\ket{c_{\text{out}}}\ket{sum} \in \{ \ket{01}, \ket{10}, \ket{11} \}$ (\cf Tab.~\ref{tab:Full-UnadderTruthTable}) are described by the sections of the circuit indicated by the separation lines, respectively.}\label{fig:Full-UnadderQCircuit}
\end{figure*}

Thus, the full-unadder can be translated to a quantum circuit acting on the initial state
\begin{equation}
    \ket{c_{\text{out}}}\ket{sum}\ket{000} \in \mathbb{C}^{2 \otimes 5}.
\end{equation}

The last $3$ qubits are then transformed into $a$, $b$, and $c_{\text{in}}$ according to Tab.~\ref{tab:Full-UnadderTruthTable}.
A quantum circuit implementing this transformation is given in Fig.~\ref{fig:Full-UnadderQCircuit}.
Filled controls are activated by $\ket{1}$, while empty controls are activated by $\ket{0}$.
The $\mathrm{R_y}$-gate denotes a rotation around the $y$-axis of the Bloch sphere describing the qubit the gate acts on. \Ie 
\begin{equation}
    \mathrm{R_y}\left(2\arccos\sqrt{\frac{2}{3}}\right) = 
    \begin{bmatrix}
        \frac{\sqrt{2}}{\sqrt{3}}   & \frac{-1}{\sqrt{3}}       \\
        \frac{1}{\sqrt{3}}          & \frac{\sqrt{2}}{\sqrt{3}}
    \end{bmatrix}.
\end{equation}

The $\mathrm{H}$- and $\mathrm{X}$-gates denote the standard Hadamard- and \texttt{NOT}-gates.
This circuit uses $5$ qubits independently of the input and is thus $\in \mathcal{O}(1)$.

\subsubsection{Optimised quantum gate for full-unadders}

It is also possible to construct a quantum gate for the single-bit full-unadder acting on only $3$ qubits instead. 
This gate then acts on the input state
\begin{equation}
    \ket{c_{\text{out}}}\ket{sum}\ket{0} \in \mathbb{C}^{2 \otimes 3}, 
\end{equation}
which are then transformed to the state 
\begin{equation}
    \ket{c_{\text{in}}}\ket{b}\ket{a} \in \mathbb{C}^{2 \otimes 3}.
\end{equation}

An explicit unitary matrix representation of such a gate is 
\begin{equation}\label{eq:OptFull-UnadderMatrix}
    \begin{bmatrix}
        1 & 0 & 0 & 0 & 0 & 0 & 0 & 0\\
        0 & \frac{1}{\sqrt{2}} & \frac{1}{\sqrt{3}} & \frac{-1}{\sqrt{6}} & 0 & 0 & 0 & 0\\
        0 & \frac{-1}{\sqrt{2}} & \frac{1}{\sqrt{3}} & \frac{-1}{\sqrt{6}} & 0 & 0 & 0 & 0\\
        0 & 0 & 0 & 0 & \frac{1}{\sqrt{3}} & \frac{1}{\sqrt{2}} & 0 & \frac{-1}{\sqrt{6}}\\
        0 & 0 & \frac{1}{\sqrt{3}} & \frac{2}{\sqrt{6}} & 0 & 0 & 0 & 0\\
        0 & 0 & 0 & 0 & \frac{1}{\sqrt{3}} & \frac{-1}{\sqrt{2}} & 0 & \frac{-1}{\sqrt{6}}\\
        0 & 0 & 0 & 0 & \frac{1}{\sqrt{3}} & 0 & 0 & \frac{2}{\sqrt{6}}\\
        0 & 0 & 0 & 0 & 0 & 0 & 1 & 0
    \end{bmatrix}.
\end{equation}

In the following, only this optimised gate is used, as it is not only enables simulation of bigger circuits due to its smaller circuit footprint, but also due to the simpler bookkeeping of the different qubits which are to be connected for more complex quantum devices. It is denoted $\mathfrak{Un}+_{\text{opt}}$ in the circuits.

\subsection{Quantum Ripple-Carry-Unadder}

Constructing a quantum \gls{rcu} with this gate is as simple as the construction of a classical \gls{rca} from full-adder circuit components.
In this construction the $c_{\text{out},0}$ is set to $\ket{0}$. It is the first input of the first full-unadder. The second input of the first unadder is the the \gls{msb} of the $n\:\qty{}{\bit}$ sum to be unadded -- \ie $sum_0$. The third qubit input of the first full-unadder is an ancillary qubit set to $\ket{0}$. The first full-unadder thus produces $\ket{c_{\text{in},0}}$ (which equals $c_{\text{out},1}$ used as the input for the second full-unadder) and the two \glspl{msb} of the $n\:\qty{}{\bit}$ numbers $a$ and $b$ -- \ie $a_0$ and $b_0$. Then, the next bit of the sum $sum_1$, $c_{\text{out},1}$, and a fresh ancilla set to $\ket{0}$ are fed into the second full-unadder. This procedure repeats until all bits of the sum $sum_i$ are fed into the device for $i \in [0,n[$. The final measurement on the corresponding qubits collapses the superposition to one of the possible input combinations satisfying Eq.~\ref{eq:RCU}.
An example of $\qty{3}{\bit}$ \gls{rcu} with input sum $6$ can be found in Appendix \ref{appendix} Fig.~\ref{fig:RCU6QCircuit}.

\subsection{Simulation of unaddition}

Now follows the description of the simulations of unaddition and their results.

\subsubsection{Simulation of full-unadder}

The simulation of the circuit from Fig.~\ref{fig:Full-UnadderQCircuit} with the \texttt{QasmSimulatorPy} simulation backend from \texttt{qiskit 0.43.3} was done with $10^9$ shots. The probabilities of a given measurement outcome are taken to be their relative frequencies. The simulated results can be seen in Tab.~\ref{tab:SimResultsFull-Unadder}.

\begin{table}[ht]
    \caption[Simulation results full-unadder]{Simulation results for the full-unadder.}
    \label{tab:SimResultsFull-Unadder}
    \centering
    \begin{tabular}{  c  c  c  c  c  c }
        \toprule
            \multicolumn{2}{c}{Input} & \multicolumn{3}{c}{Output}  & Probability $[\qty{}{\percent}]$ \\
            $c_{\text{out}}$  & $sum$ & $c_{\text{in}}$ & $b$ & $a$ & \\
        \midrule
            $0$               & $0$   & $0$ & $0$ & $0$ & \qty{100,0000000}{} \\
        \midrule
                              &       & $1$ & $0$ & $0$ & \qty{ 33,3326319}{} \\
            $0$               & $1$   & $0$ & $1$ & $0$ & \qty{ 33,3335182}{} \\
                              &       & $0$ & $0$ & $1$ & \qty{ 33,3338499}{} \\
        \midrule
                              &       & $0$ & $1$ & $1$ & \qty{ 33,3341311}{} \\
            $1$               & $0$   & $1$ & $0$ & $1$ & \qty{ 33,3335476}{} \\
                              &       & $1$ & $1$ & $0$ & \qty{ 33,3323213}{} \\
        \midrule
            $1$               & $1$   & $1$ & $1$ & $1$ & \qty{100,0000000}{} \\
        \botrule
    \end{tabular}
    \footnotetext{Simulation results for the full-unadder quantum circuit (\cf Fig.~\ref{fig:Full-UnadderQCircuit}) for all possible input bit combinations. The simulation was done for \qty{e9}{} shots and the probabilities were calculated as the relative frequency of a given output.}
\end{table}

Similarly, the simulation of the optimised gate version of the full-unadder given by Eq.~\ref{eq:OptFull-UnadderMatrix} was done with the \texttt{AerSimulator} backend from \texttt{qiskit 0.43.3} for $10^9$ shots with probabilities of each outcome taken to be their relative frequencies again. These result is presented in Tab.~\ref{tab:SimResultsOptFull-Unadder}.
The comparison of these results quite clearly show the identical behaviour for the two devices within a reasonable margin of error.

\begin{table}[ht]
    \caption[Simulation results optimised full-unadder]{Simulation results for the optimised full-unadder.}
    \label{tab:SimResultsOptFull-Unadder}
    \centering
    \begin{tabular}{  c  c  c  c  c  c }
        \toprule
            \multicolumn{2}{c}{Input} & \multicolumn{3}{c}{Output}  & Probability $[\qty{}{\percent}]$ \\
            $c_{\text{out}}$  & $sum$ & $c_{\text{in}}$ & $b$ & $a$ & \\
        \midrule
            $0$               & $0$   & $0$ & $0$ & $0$ & \qty{100,0000000}{} \\
        \midrule
                              &       & $1$ & $0$ & $0$ & \qty{ 33,3342689}{} \\
            $0$               & $1$   & $0$ & $1$ & $0$ & \qty{ 33,3310289}{} \\
                              &       & $0$ & $0$ & $1$ & \qty{ 33,3347022}{} \\
        \midrule
                              &       & $0$ & $1$ & $1$ & \qty{ 33,3365392}{} \\
            $1$               & $0$   & $1$ & $0$ & $1$ & \qty{ 33,3322202}{} \\
                              &       & $1$ & $1$ & $0$ & \qty{ 33,3312406}{} \\
        \midrule
            $1$               & $1$   & $1$ & $1$ & $1$ & \qty{100,0000000}{} \\
        \botrule
    \end{tabular}
    \footnotetext{Simulation results for the optimised full-unadder quantum gate (\cf Eq.~\ref{eq:OptFull-UnadderMatrix}) for all possible input bit combinations. The simulation was done for \qty{e9}{} shots and the probabilities were calculated as the relative frequency of a given output.}
\end{table}

\subsubsection{Simulation of Ripple-Carry-Unadder}

The transpiled circuit if the \gls{rcu} using the optimised full-unadder gate was simulated with the \texttt{AerSimulator} backend and \texttt{qiskit 0.43.3}, because the \texttt{QasmSimulatorPy} backend is limited to maximally $24$ qubits. This enables the simulation of bigger unadditions. Thus, it was possible to to simulate the unaddition of input number $N$ in the range \qtyrange{0}{524287}{}, that is, up to $2^{19} -1$ on the available classical hardware.
The simulations results in the set of the triples $(a,b,c_{\text{in}})$ for $n\:\qty{}{\bit}$ $a$, $b$, and \qty{1}{\bit} $c_{\text{in}}$ according to Eq.~\ref{eq:RCU}, provided that the simulation was done with a sufficient number of shots to observe all $2N+1$ expected triples. 
This confirms with very solid data basis that the \gls{rcu} works as expected.

\subsection{Unmultiplication}\label{subsec:Unmultiplication}

\begin{figure}[!bt]
    \centering
    \includegraphics[width=0.5\textwidth]{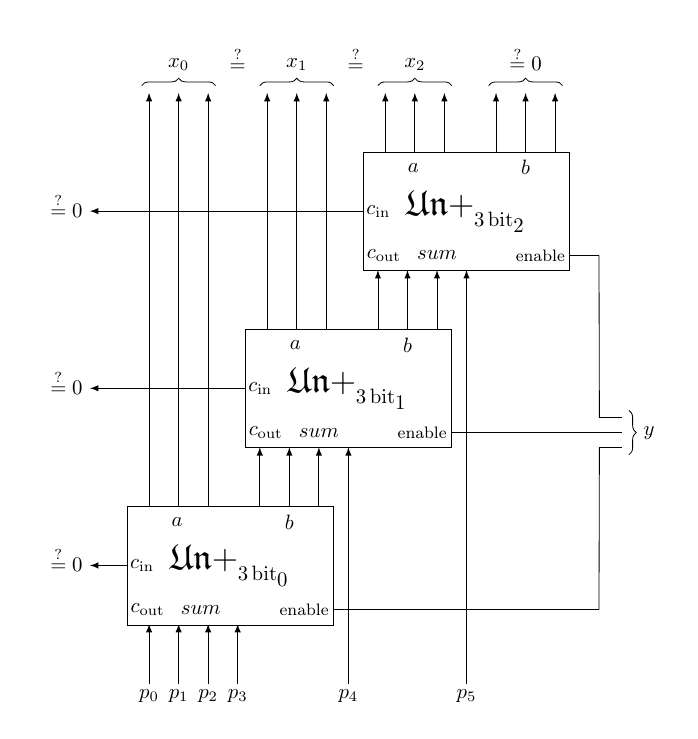}
    \caption[Black-box circuit of a \qty{3}{\bit} unmultiplier]{Black box circuit of a \qty{3}{\bit} unmultiplier. It is the unoperation device corresponding to the classical circuit for the multiplier built from \glspl{rca} and in turn full-adders (\cf \cite[Example 12.21]{IntroLogicCircuits-ArithmeticCircuits}). It is capable of the unmultiplication of numbers whose factors fit into the \qty{3}{\bit} registers. The "enable" describes the binary multiplication with $1$ or $0$, \ie if the enable signal is $0$ ($1$) the corresponding \gls{rcu} is off (on).}
    \label{fig:UnmultBlackBox}
\end{figure}

The preceding elaborations are a proof of principle, yet the unoperation provided for addition does not really solve any computationally interesting problem since both directions of summation are computationally unproblematic. The problem of factoring integer numbers on the other hand is considered hard and utilisable for computational security \cite{RSA, DiffieHellman}, which is still true classically \cite{NumberFieldSieve}, albeit not quantum \cite{ShorsAlgorithm, ShorsAlgorithm94}. This problem can also be approached by the unoperation of multiplication which will return the set of all factor pairs that can be multiplied to get the input number. The unmultiplication is denoted $\mathfrak{Un}\times$. The construction of a corresponding device again utilises the \textrm{divide et impera} approach which breaks down classical multiplication into a series of additions \cite[pp.405 et seqq.]{IntroLogicCircuits-ArithmeticCircuits}. The unmultiplication device now reverses the information flow and, consequently, exchanges the \glspl{rca} with \glspl{rcu}. 
The black box construction of such a device can be seen in Fig.~\ref{fig:UnmultBlackBox}.

First, let us look at the classical binary multiplier that acts on inputs $x$ and $y$ to produce product $p$, in our example both $x$ and $y$ fit into \qty{3}{\bit} and thus $p$ fits into \qty{6}{bit}. 
It would then take the input $x$ and add it onto an empty register conditioned on $y$ -- \ie if the \gls{lsb} of $y$ is $0$ the register containing the result $p$ remains empty, if it is $1$ then $x$ is added, this \qty{3}{\bit} addition of course having $c_{\text{in}} = 0$ for its \gls{lsb}-addition. The output of this first addition is a \qty{4}{\bit} number, where $c_{\text{out}}$ houses the \gls{msb} and the remaining bits are saved in sum. The same is then performed for the remaining bits in $y$ with the input of the next \gls{rca} being the 3 \glspl{msb} of the previous addition, while its \gls{lsb} is fixed as a resulting bit in the register of $p$.
\cite[pp.405 et seqq.]{IntroLogicCircuits-ArithmeticCircuits}

Because the multiplier of such a construction just employs controlled additions, it is possible to unoperate it using the unadder as follows.
The black-box construction of the unmultiplier reverses the flow of information to construct a superposition of all possible inputs to the corresponding classical binary multiplier with the analogous construction described above. This can be seen in Fig.~\ref{fig:UnmultBlackBox} for the \qty{3}{\bit} unmultiplier. It takes the 4 \glspl{msb} $(p_0, p_1, p_2, p_3)$ of product $p$ as inputs to the first \qty{3}{\bit} \gls{rcu} (constructed as in Fig.~\ref{fig:RCU6QCircuit}). Conditioned on the output of this \gls{rcu} the \gls{msb} of $y$ is reconstructed. Refer to the Sec.~\ref{subsec:UnmultQuantumCircuit} for how this is done specifically.

Outputs $a$ (as register $x_1$) and $c_{\text{in}}$ of the first \gls{rcu} are used for later classical post-processing, while $b$ (\qty{3}{\bit}) is fed into the second \gls{rcu} as the 3 \glspl{msb} of the sum input and the next bit of the product ($p_4$) is used as the \gls{lsb}. The outputs $c_{\text{in}}$ and $a$ (as register $x_2$) of this \gls{rcu} are again used in post-processing and $b$ as the 3 \glspl{msb} for the third \gls{rcu}. The procedure is repeated for the third -- and last -- \gls{rcu}, the only difference being that its $b$ output is not fed into further devices but is also used for post-processing (\cf Sec.~\ref{subsec:UnmultQuantumCircuit}).

Accordingly, for an $n\:\qty{}{\bit}$ unmultiplier this procedure would be repeated with $n$ separate $n\:$\qty{}{\bit} \glspl{rcu}.

\subsection{Quantum circuit for unmultiplication}\label{subsec:UnmultQuantumCircuit}

It is relatively straightforward to translate the black-box view (Fig.~\ref{fig:UnmultBlackBox}) to a quantum circuit. The corresponding quantum circuit that implements the unmultiplication for \qty{3}{\bit} with the described construction is shown in Appendix \ref{appendix} Fig.~\ref{fig:UnmultQCircuit} for an input product $p = 6$ (binary $000110$).
The register $p$ is fed to the \qty{3}{\bit} \gls{rcu}-gates as described in Sec.~\ref{subsec:Unmultiplication}, while registers $x_i$ with $i \in \{0,1,2\}$ are used as the ancilla inputs which are transformed to the $a$-outputs of the corresponding \gls{rcu} $\mathfrak{Un}+_{\text{opt},3\:\mathrm{bit}_i}$.

The reconstruction of the factor $y$ of the product $p = x \cdot y$ is done with the procedure termed \gls{qfb} for $y$. This $y$-\gls{qfb} means, that the qubit of $y$ corresponding to $\mathfrak{Un}+_{\text{opt},3\:\mathrm{bit}_{i}}$ ($y_i$) is set to $1$ with the $\mathrm{X}$-gate and then activates the \gls{rcu}. After the \gls{rcu}, if its $a$-output ($x_i$) is $0$ on all bits, $y_i$ is set to $0$, as this means that the \gls{rcu} did not act, which is the equivalent of the conditional enable signal in the classical binary multiplier being $0$. This is implemented with the multi-$0$-controlled $\mathrm{X}$-gate as shown in Fig.~\ref{fig:UnmultQCircuit} after each \gls{rcu}-gate.

After the 3 \gls{rcu}- and $y$-\gls{qfb} procedures on the relevant qubits the state is measured to the corresponding classical registers $c_{\text{in},x_i}$, $x_i$, $y$ and $const0$ and post-processing can begin.

Because there is the initial state of the classical binary multiplier which is fixed for all inputs -- that is, it uses $0$ as all $c_{\text{in}}$ of all \glspl{rca}, $0$ in the $b$-register of the first \gls{rca}, and it has only a single $x$ which is the $a$-input for all \glspl{rca} -- the unmultiplier has to take care of this. 
The unmultiplier has outputs which will be part of the superposition constructed by the device, yet are not valid inputs to a classical binary multiplier, \eg some $c_{\text{in}} \neq 0$, the $b$-output of the last \gls{rcu} $\neq 0$ or multiple different $x_i$. These have to be sorted out by classical post-processing of the measured outcomes after the quantum circuit.
\Ie on the observed outcomes one has to perform the classical checks if 
\begin{align}
    const0 &\overset{?}{=} 0, \\
    c_{\text{in}, x_i} &\overset{?}{=} 0 \:\:\forall\: i, \\
    (x_i \overset{?}{=} x_j) &\lor (x_i \overset{?}{=} 0) \:\:\forall\: i \neq j.
\end{align}

After that we get all possible factor pairs $(x, y)$ that produce $p = x \cdot y$.

With the construction from Fig.~\ref{fig:UnmultQCircuit} generalised to $n\:$\qty{}{\bit} on the available classical hardware it was possible to simulate the inputs $p \in [0,16[$. 
For all simulated products the unmultiplier produces all possible factor pairs that produce the given $p$.

Regarding complexity, this unmultiplier construction for $n\:\qty{}{\bit}$ input number $N$ using the optimised full-unadder gate is 
\begin{equation*}
    \in \mathcal{O}(n^2 + 3n) \subseteq \mathcal{O}(n^2) = \mathcal{O}(\lceil\log_2{N}\rceil^2)
\end{equation*}
in the number of qubits. Asymptotically, one can omit the rounding $\lceil\cdot\rceil$ and the base of the logarithm, thus this unmultiplier is $\in \mathcal{O}((\log{N})^2)$.

\section{Discussion}\label{sec:Discussion}

In conclusion, this work shows that it is possible at least for some operations to construct the corresponding unoperation within the quantum circuit framework, which may be utilised to build an efficient device for factoring. 
It not only shows the existence of such an algorithm, but provides a constructive approach to implement such an algorithm for the unoperation of addition and multiplication.
The proposed devices are easily implemented and simulatable on reasonably powerful classical hardware for nontrivial examples to show their working principle.

Aside from presented optimisations, other schemes for improving the performance and the resource requirements can be implemented. These could include techniques to modify the devices in order to decrease the amount of possible outcomes to be post-processed classically that do not correspond to factor pairs directly, or to recover factor pairs from such an outcome. 

Also different different classical multiplication algorithms can be explored for applicability of the unoperation scheme
Even other trapdoor functions where no efficient algorithm for computing their reverse direction may be unoperated, opening up the possibility for different quantum cryptanalytic algorithms.

\backmatter

\bmhead{Acknowledgements}

The author acknowledges the financial support by the Federal Ministry of Research, Technology and Space of Germany (BMFTR) via the Q-TREX project with identification number 16KISR026.

\section*{Declarations}

\subsection*{Funding}

PK: Federal Ministry of Research, Technology and Space of Germany (BMFTR), 16KISR026.

\subsection*{Data availability}

All data that support the findings of this study are available from the author upon reasonable request.

\subsection*{Code availability}

The code for simulations is available from the author upon reasonable request.

\subsection*{Author contributions}

PK wrote all sections, constructed all figures and tables, wrote and performed simulations, and reviewed the manuscript.

\subsection*{Competing Interests}

The author declares no conflict of interest.

\begin{appendix} 

\section{Quantum circuits}\label{appendix}

\begin{figure*}[htb]
    \phantomsection
    \centering
    \includegraphics[width=1.0\textwidth]{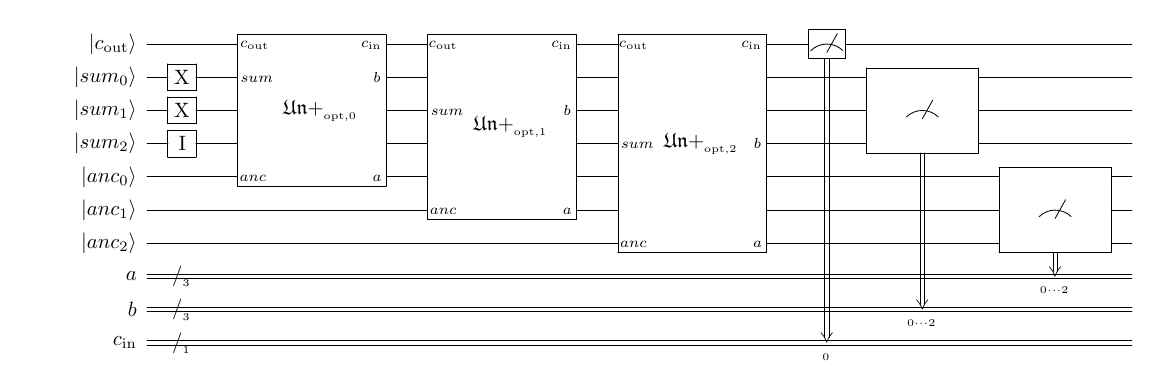}
    \caption[Quantum circuit of a 3 bit RCU]{Quantum circuit of a \qty{3}{\bit} \gls{rcu} with input sum $6$ (binary $110$). It uses the optimised full-unadder gate $\mathfrak{Un}+_{\text{opt}}$ implemented with the unitary matrix from Eq.~\ref{eq:OptFull-UnadderMatrix}.}
    \label{fig:RCU6QCircuit}
\end{figure*}

\begin{figure*}[phtb]
    \phantomsection
    \centering
    \includegraphics[angle=90, width=0.6\textwidth]{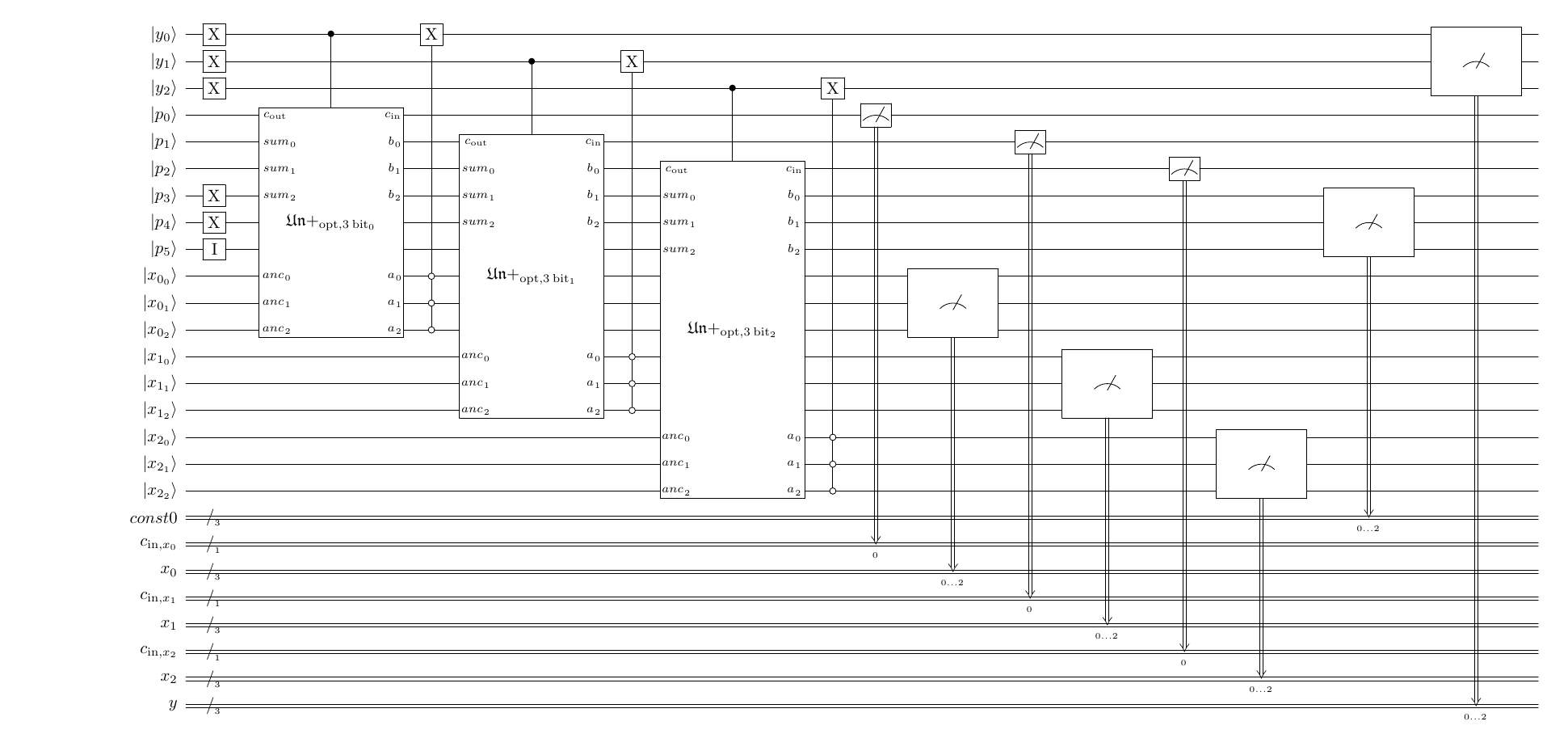}
    \caption[Quantum circuit of a 3 bit Unmultiplier]{Quantum circuit of a \qty{3}{\bit} unmultiplier with input product $6$ (binary $000110$, leading zeroes implicit in the circuit for clarity) with \gls{qfb} to recover factor $y$. It uses \qty{3}{\bit} \glspl{rcu} denoted $\mathfrak{Un}+_{\text{opt},3\:\mathrm{bit}_i}$ for $i \in \{0,1,2\}$, which are implemented with optimised full-unadder gate $\mathfrak{Un}+_{\text{opt}}$ analogously to Fig.~\ref{fig:RCU6QCircuit}.}
    \label{fig:UnmultQCircuit}
\end{figure*}

The quantum circuit of the $\qty{3}{\bit}$ \gls{rcu} with input $6$ can be found in Fig.~\ref{fig:RCU6QCircuit}.
The unmultiplier circuit can be found in Fig.~\ref{fig:UnmultQCircuit}.

\end{appendix} 

\printbibliography[title=References]

@article{ CommentOnQuantumIdentityAuthenticationWithSinglePhoton,
    title = {Comment on “quantum identity authentication with single photon”},
    journal = {Quantum Information Processing},
    volume = {23},
    pages = {357},
    number = {10},
    year = {2024},
    month = {10},
    day = {19},
    issn = {1573-1332},
    doi = {10.1007/s11128-024-04564-x},
    author = {Davide {Li Calsi} and Paul Kohl}
}

@article{ BB84Protocol,
    title = {Quantum cryptography: Public key distribution and coin tossing},
    journal = {Theoretical Computer Science},
    volume = {560},
    pages = {7-11},
    year = {2014},
    note = {Theoretical Aspects of Quantum Cryptography – celebrating 30 years of BB84},
    issn = {0304-3975},
    doi = {10.1016/j.tcs.2014.05.025},
    author = {Charles H. Bennett and Gilles Brassard}
}

@article{ DiffieHellman,
    author={Whitfield Diffie and Martin E. Hellman},
    journal={IEEE Transactions on Information Theory},
    title={New Directions in Cryptography},
    year={1976},
    volume={22},
    number={6},
    pages={644-654},
    doi={10.1109/TIT.1976.1055638},
    ISSN={1557-9654},
    month={11},
}

@article{ E91Protocol,
    title = {Quantum cryptography based on Bell's theorem},
    author = {Ekert, Artur K.},
    journal = {Phys. Rev. Lett.},
    volume = {67},
    issue = {6},
    pages = {661--663},
    numpages = {3},
    year = {1991},
    month = {08},
    publisher = {American Physical Society},
    doi = {10.1103/PhysRevLett.67.661},
}

@inproceedings{ GroversAlgorithm,
    title={A fast quantum mechanical algorithm for database search},
    author={Lov K. Grover},
    year={1996},
    isbn = {978-0-89791-785-8},
    publisher = {Association for Computing Machinery (ACM)},
    doi = {10.1145/237814.237866},
    booktitle = {Proceedings of the Twenty-Eighth Annual ACM Symposium on Theory of Computing},
    pages = {212–219},
    numpages = {8},
    location = {Philadelphia, Pennsylvania, USA}, 
    series = {STOC '96},
    eprint={quant-ph/9605043v3},
    archivePrefix={arXiv},
    primaryClass={quant-ph},
}

@Inbook{ IntroLogicCircuits-ArithmeticCircuits,
    author = {LaMeres, Brock J.},
    title = {Arithmetic Circuits},
    bookTitle = {Introduction to Logic Circuits {\&} Logic Design with Verilog},
    year = {2023},
    adress = {Cham},
    edition = {3},
    publisher = {Springer International Publishing},
    pages = {389--418},
    isbn = {978-3-031-43946-9},
    doi = {10.1007/978-3-031-43946-9_12},
}

@inproceedings{ NumberFieldSieve,
    author = {Lenstra, A. K. and Lenstra, H. W. and Manasse, M. S. and Pollard, J. M.},
    title = {The Number Field Sieve},
    year = {1990},
    isbn = {978-0-89791-361-4},
    publisher = {Association for Computing Machinery},
    address = {New York, NY, USA},
    doi = {10.1145/100216.100295},
    booktitle = {Proceedings of the Twenty-Second Annual ACM Symposium on Theory of Computing},
    pages = {564–572},
    numpages = {9},
    location = {Baltimore, Maryland, USA},
    series = {STOC ’90},
}

@misc{ Qiskit,
    title={Quantum computing with Qiskit}, 
    author={Ali Javadi-Abhari and Matthew Treinish and Kevin Krsulich and Christopher J. Wood and Jake Lishman and Julien Gacon and Simon Martiel and Paul D. Nation and Lev S. Bishop and Andrew W. Cross and Blake R. Johnson and Jay M. Gambetta},
    year={2024},
    eprint={2405.08810v3},
    archivePrefix={arXiv},
    primaryClass={quant-ph},
    doi={10.48550/arXiv.2405.08810},
}

@software{ QiskitSoftwarePackage,
    author       = {Gadi Aleksandrowicz and
                  Thomas Alexander and
                  Panagiotis Barkoutsos and
                  Luciano Bello and
                  Yael Ben-Haim and
                  David Bucher and
                  Francisco Jose Cabrera-Hernández and
                  Jorge Carballo-Franquis and
                  Adrian Chen and
                  Chun-Fu Chen and
                  Jerry M. Chow and
                  Antonio D. Córcoles-Gonzales and
                  Abigail J. Cross and
                  Andrew Cross and
                  Juan Cruz-Benito and
                  Chris Culver and
                  Salvador De La Puente González and
                  Enrique De La Torre and
                  Delton Ding and
                  Eugene Dumitrescu and
                  Ivan Duran and
                  Pieter Eendebak and
                  Mark Everitt and
                  Ismael Faro Sertage and
                  Albert Frisch and
                  Andreas Fuhrer and
                  Jay Gambetta and
                  Borja Godoy Gago and
                  Juan Gomez-Mosquera and
                  Donny Greenberg and
                  Ikko Hamamura and
                  Vojtech Havlicek and
                  Joe Hellmers and
                  Łukasz Herok and
                  Hiroshi Horii and
                  Shaohan Hu and
                  Takashi Imamichi and
                  Toshinari Itoko and
                  Ali Javadi-Abhari and
                  Naoki Kanazawa and
                  Anton Karazeev and
                  Kevin Krsulich and
                  Peng Liu and
                  Yang Luh and
                  Yunho Maeng and
                  Manoel Marques and
                  Francisco Jose Martín-Fernández and
                  Douglas T. McClure and
                  David McKay and
                  Srujan Meesala and
                  Antonio Mezzacapo and
                  Nikolaj Moll and
                  Diego Moreda Rodríguez and
                  Giacomo Nannicini and
                  Paul Nation and
                  Pauline Ollitrault and
                  Lee James O'Riordan and
                  Hanhee Paik and
                  Jesús Pérez and
                  Anna Phan and
                  Marco Pistoia and
                  Viktor Prutyanov and
                  Max Reuter and
                  Julia Rice and
                  Abdón Rodríguez Davila and
                  Raymond Harry Putra Rudy and
                  Mingi Ryu and
                  Ninad Sathaye and
                  Chris Schnabel and
                  Eddie Schoute and
                  Kanav Setia and
                  Yunong Shi and
                  Adenilton Silva and
                  Yukio Siraichi and
                  Seyon Sivarajah and
                  John A. Smolin and
                  Mathias Soeken and
                  Hitomi Takahashi and
                  Ivano Tavernelli and
                  Charles Taylor and
                  Pete Taylour and
                  Kenso Trabing and
                  Matthew Treinish and
                  Wes Turner and
                  Desiree Vogt-Lee and
                  Christophe Vuillot and
                  Jonathan A. Wildstrom and
                  Jessica Wilson and
                  Erick Winston and
                  Christopher Wood and
                  Stephen Wood and
                  Stefan Wörner and
                  Ismail Yunus Akhalwaya and
                  Christa Zoufal},
    title        = {Qiskit: An Open-source Framework for Quantum
                   Computing
                  },
    month        = {1},
    year         = {2019},
    publisher    = {Zenodo},
    version      = {0.7.2},
    doi          = {10.5281/zenodo.2562111},
}

@article{ QuantumAuthShi,
    title = {Quantum key distribution and quantum authentication based on entangled state},
    journal = {Physics Letters A},
    publisher={Elsevier},
    volume = {281},
    number = {2},
    pages = {83-87},
    year = {2001},
    number={2–3},
    month= {3}, 
    issn = {0375-9601},
    author = {Bao-Sen Shi and Jian Li and Jin-Ming Liu and Xiao-Feng Fan and Guang-Can Guo},
    DOI={10.1016/s0375-9601(01)00129-3},
    eprint={quant-ph/0102058v1},
    archivePrefix={arXiv},
    primaryClass={quant-ph},
}

@article{ QuantumAuthShiCommentWei,
    title={Comment on “Quantum Key Distribution and Quantum Authentication Based on Entangled State”},
    author={Wei, Toung-Shang and Tsai, Chia-Wei and Hwang, Tzonelih},
    journal={International Journal of Theoretical Physics},
    volume={50},
    issue={9},
    pages={2703--2707},
    year={2011},
    publisher={Springer},
    issn={1572-9575},
    doi={10.1007/s10773-011-0768-0}
}

@article{ QuantumMACCurty,
    title = {Quantum authentication of classical messages},
    author = {Curty, Marcos and Santos, David J.},
    journal={Physical Review A},
    volume = {64},
    issue = {6},
    pages = {062309},
    number= {6},
    year = {2001},
    month = {11},
    publisher = {American Physical Society},
    doi = {10.1103/PhysRevA.64.062309},
    ISSN={1094-1622},
    eprint={quant-ph/0103122v2},
    archivePrefix={arXiv},
    primaryClass={quant-ph},
}

@article{ QuantumIdentityAuthHong,
    title = {Quantum identity authentication with single photon},
    author = {Hong, {Chang Ho} and Heo, Jino and Jang, {Jin Gak} and Kwon, Daesung},
    journal={Quantum Information Processing},
    volume = {16},
    issue = {10},
    pages = {236},
    year = {2017},
    month = {08},
    doi = {10.1007/s11128-017-1681-0},
    ISSN={1573-1332},
}

@article{ QuantumIdentityAuthZawadzki,
    title = {Quantum identity authentication without entanglement},
    author = {Zawadzki, Piotr},
    journal={Quantum Information Processing},
    volume = {18},
    issue = {1},
    pages = {7},
    year = {2018},
    month = {11},
    doi = {10.1007/s11128-018-2124-2},
    ISSN={1573-1332},
}

@Article{ QuantumIdentityAuthZawadzkiAttackGonzalez,
    AUTHOR = {González-Guillén, Carlos E. and González Vasco, María Isabel and Johnson, Floyd and Pérez del Pozo, Ángel L.},
    TITLE = {An Attack on Zawadzki’s Quantum Authentication Scheme},
    JOURNAL = {Entropy},
    VOLUME = {23},
    YEAR = {2021},
    NUMBER = {4},
    ARTICLE-NUMBER = {389},
    PubMedID = {33805925},
    ISSN = {1099-4300},
    DOI = {10.3390/e23040389},
}

@article{ RSA,
    author = {Ron L. Rivest and Adi Shamir and Leonard Adleman},
    title = {A method for obtaining digital signatures and public-key cryptosystems},
    journal = {Communications of the ACM},
    year = {1978},
    month = {2},
    publisher = {Association for Computing Machinery},
    volume = {21},
    number = {2},
    issn = {0001-0782},
    doi = {10.1145/359340.359342},
    pages = {120–126},
    numpages = {7},
}

@article{ ShorsAlgorithm,
    title={Polynomial-Time Algorithms for Prime Factorization and Discrete Logarithms on a Quantum Computer},
    volume={26},
    ISSN={1095-7111},
    DOI={10.1137/s0097539795293172},
    number={5},
    journal={SIAM Journal on Computing},
    publisher={Society for Industrial & Applied Mathematics (SIAM)},
    author={Peter W. Shor},
    year={1997},
    month={10},
    pages={1484–1509}
}

@inproceedings{ ShorsAlgorithm94,
    author={Peter W. Shor},
    booktitle={Proceedings 35th Annual Symposium on Foundations of Computer Science},
    title={Algorithms for quantum computation: discrete logarithms and factoring},
    year={1994},
    pages={124-134},
    doi={10.1109/SFCS.1994.365700},
}

@book{ TechnischeInformatik1GrundlagenDerDigitalenElektronik,
    title =     {Technische Informatik 1: Grundlagen der digitalen Elektronik},
    author =    {Wolfram Schiffmann and Robert Schmitz},
    publisher = {Springer},
    isbn =      {978-3-642-18894-7},
    doi =       {10.1007/978-3-642-18894-7},
    year =      {2004},
    series =    {Springer-Lehrbuch},
    edition =   {5},
}

@article{ VariationalQuantumEigensolver,
    title={A variational eigenvalue solver on a photonic quantum processor},
    volume={5},
    ISSN={2041-1723},
    DOI={10.1038/ncomms5213},
    number={1},
    journal={Nature Communications},
    publisher={Springer Science and Business Media LLC},
    author={Peruzzo, Alberto and McClean, Jarrod and Shadbolt, Peter and Yung, Man-Hong and Zhou, Xiao-Qi and Love, Peter J. and Aspuru-Guzik, Alán and O’Brien, Jeremy L.},
    year={2014},
    month={7},
    pages={4213},
    eprint={1304.3061v1},
    archivePrefix={arXiv},
    primaryClass={quant-ph},
}

\end{document}